\documentclass[10pt]{article}
\usepackage{amsfonts}
\usepackage{amsmath}
\usepackage{graphicx}
\usepackage{epsfig}
\usepackage{color}
\usepackage{textcomp}

\begin{document}

\title{Discrete Opinion Dynamics with $M$ choices}
\author{Andr\'e C. R. Martins\\
NISC -- EACH -- Universidade de S\~ao Paulo,\\
Rua Arlindo B\'etio, 1000, 03828--000,  S\~ao Paulo, Brazil}

\date{amartins@usp.br}

\maketitle

\begin{abstract}

Here, I study how to obtain an opinion dynamics model for the case where there are $M$ possible discrete choices and there is need to model how strong each agent choice is. The new model is obtained as an extension of the Continuous Opinions and Discrete Actions (CODA) model. Technical difficulties with the choice of proper variables for a simple model are solved. For the symmetrical case, a dimensionless model is found. However, when analyzing the results, a change of variables seems to be required for ease of interpretation. Extremism is observed here as well, generated by the local reinforcement of opinions inside domains of agents with the same choice.

\end{abstract}

%Key words: Sociophysics, opinion dynamics, CODA, GUF

\section{Introduction}

 Most opinion Dynamics ~\cite{castellanoetal07,galametal82,galammoscovici91,sznajd00,deffuantetal00,deffuantetal02a,amblarddeffuant04,galam05,weisbuchetal05,franksetal08a,martins08a,martins12b,Li2013,Parsegov2017,Amelkin2017,Sirbu2017} either focus on two choices or define opinion as a value over a continuous range. However there are several real life situations when people face a choice over more than two choices. If those choices correspond to an ideological position that can be represented as projections over a one-dimensional direction, continuous opinion models could, in principle, be used as an approximation to several choices if ranges are defined over the possible values. By making the role of choosing clearer, a model with actual many choices might be an improvement over that approximation. Still, as all the available models are already approximations, using continuous models we already know could be a good way to move forward. However, when all possible choices are independent and there is no reasonable one-dimensional projection, that strategy makes no sense. Examples of situations where those independent choices happen are not few: choosing a brand when there are several available, or a sport team, or even picking a candidate among many in an election not heavily influenced by a single ideological issue. In those situations, we need models that allow for any number $M$ of choices.
 
 %by introducing the notion of inflexibles. It was shown, by using GUF \cite{galam05b}, that depending on the proportion of inflexibles at each side of a debate, even a small minority could eventually convince the whole population about their ideas.

 Models with three possible choices do exist and they have become less rare more recently~\cite{galam90a,galam91a,vazquezredner04a,gekleetal05a,delaLama2006,martins10b,Galam2013,Wu2014,Wu2018}. Sometimes, those three options do represent independent choices, but the case where there is a third intermediary option, be it undecided agents or centrists has also been explored. Strongly opinionated individuals can play a very important role in a public debate, even when debate only happens in small groups. By simply holding to their opinions when meeting different groups of people, they can cause a minority to slowly increase in size until it becomes a majority and wins the debate \cite{galam05,galamjacobs07,Couzin2011,Balenzuela2015}.

 Here, I will extend the Continuous Opinions and Discrete Action (CODA) model \cite{martins08a,martins12b} to deal with any number $M$ of choices, taking it beyond a previous study of the $M=3$ case~\cite{martins10b}. The general model presented here can be used for any type of choices, be them independent, aligned on a number of ideological dimensions or any mixture of those. The general case will be explained I will show how different assumptions constrain the likelihood matrix parameters of how agents influence each other. In particular, in the symmetric case, we will see that the likelihood matrix has only one independent value for any value of $M$ and if we look only at the dynamics of opinions, even that value is irrelevant, as is the case in the traditional CODA. The one-dimensional ideological case will also be debated and we will see that it does not lead naturally to such an elegant scenario as in the symmetrical situation. Instead, most of the constraints we get are inequalities between the likelihood parameters. That is, for large $M$ we will still have many independent parameters that only have to obey ranking relationships. Even assuming a reflection symmetry around the middle opinions does not make the model to have few parameters for large $M$s.
 
 To explore the new model in the simpler symmetrical case, simulations have been prepared. Their results highlight how extreme opinions can be also when we have more choices. We will also see the consequences of having more choices for the size of the agreement clusters. Analyzing the simulation outcomes also helps at understanding a technical problem that comes from this natural extension of CODA. The simpler model is obtained when we use the odds between sequential two choices to obtain. But that choice of variables is not so easy to interpret. That problem can be solved if we keep the simplified model for the simulation but interpret the output using a transformation to easier variables.

\section{The model}

Following the notation introduced in a previous paper on the use of Bayesian modeling as a framework for obtaining opinion dynamics models\cite{martins12b}, each agent $i$ has a personal choice among the values $x_m$, where $x_m = 1, \ldots, M$ is one of $M$ possible finite values $1,\cdots , M$, representing each one of the possible $M$ choices. For each value of $x_m$, agent $i$ has a probabilistic opinion $f_i(x_m)$ that measures $i$ belief about how probable it is that each choice is the best one. In the general case, when interacting with other agents (or observed by them), the actual choice other agents observe is obtained as a functional of the probability distribution $f_i$, $a_i=A[f_i]$. Here, we will simply assume that this is determined by the case where the value of $f_i(x)$ is maximum.

Each agent assumes there is a best choice $x^*$ that is not known, only inferred by its neighbors. When updating its opinion based on the observation of a neighbor $j$, agent $i$ assumes, given each possible choice $x^*$, there is a likelihood $p(a_m|x^*)$ that a neighbor will choose each possible $a_m$. This can be more easily understood and represented as a likelihood matrix $\mathbb{L}_{mn}=p(a_j=m|x^*=n)$, where, obviously, $\sum_m \mathbb{L}_{mn}$ must be 1 for all $n$. That means, in the general case, we have $M^2-M$ independent values for the likelihood matrix. That assumes all agents have homogeneous beliefs about all neighbors. A more general case can be obtained where each agent $i$ has a different likelihood matrices for each neighbor $j$. That would be represented by a matrix  $\mathbb{L}_{mnij}$. This case will not be discussed here but it can be useful to represent different characteristics, such as the existence of contrarians \cite{galam04,delalamaetal05a,martinskuba09a,hongstrogatz11a} or the evolution of trust  \cite{boeroetal10a,carlettietal10a,sietal10a,sietal10b,richterspeixoto11a,fanetal21a} between the agents\cite{martins13b}.

Given the initial probabilistic opinion $f_i(x,t)$ agent $i$ has about each possible value of $x_n$ at instant $t$ and that agent $i$ observes its neighbor $j$ favors a choice $a_m$, an update rule can be trivially obtained by applying Bayes Theorem to the problem. We have, for each possible value $x_n$, assuming agent $j$ was observed with choice $a_m$, that
\begin{equation}\label{eq:bayes}
f_i(x_n,t+1)= \frac{1}{N} f_i(x_n,t)\mathbb{L}_{mn},
\end{equation}
where $N=\sum_o f_i(x_o,t)\mathbb{L}_{mo}$ is a normalizing constant needed to ensure that $\sum_n f_i(x_n,t+1)$ adds up to 1.

In the case of two choices $a_1$ and $a_2$, that is, $M=2$, it was possible to greatly simplify the model by noticing that $p(t)=f(x_1,t)= 1-f(x_2,t)$ and reducing the problem to the logodds variable
\[
\nu_i(t)=\ln{\frac{p_i(t)}{1-p_i(t)}},
\] 
where the choice of agent $i$ is easily obtained from the sign of $\nu_i$. The division cancels the normalization constant and, assuming agent $j$ chooses $a_1$ the logarithm turns the model into the simple additive rule
\begin{equation}\label{eq:coda}
\nu_i(t+1)=\nu_i(t)+\left[\ln(\mathbb{L}_{11})- \ln(\mathbb{L}_{21}) \right],
\end{equation}
where the term in brackets is constant. That allows a renormalized version of the equation \cite{martins08b} where, in the general case, we simply add $\pm 1$, depending on the choice of the neighbor.

\subsection{Choosing convenient variables}

In the general case of $M$ possible choices, we can not reduce the probability problem to one variable. Instead, each agent has $M-1$ independent values $f_i(x_m)$ that must be updated with each interaction between two agents. If we still want to get rid of the normalization constant and obtain an additive model, we have now two possible choices for defining the variable we will actually use. One possible attempt to do that is to define a log-odds opinion of agent $i$ about choice $m$, given by
\[
\nu_{in}(t)=\ln{\frac{ f_i(x_n,t)(t)}{1- f_i(x_n,t)(t)}}.
\] 
This choice, however, produces a not so simple dynamics. When updating $\nu_{im}(t+1)$ from the observation that agent $j$ prefers $m$, we no longer obtain the simplifications that led to Equation \ref{eq:coda}. Here, $1- f_i(x_n,t)(t)$ must be calculated as the sum of all other terms, that is, we need to update the probabilities in
\[
\nu_{in}(t+1)=\ln{\frac{f_i(x_n,t)(t+1)}{\sum_{m\neq n}f_i(x_m,t)(t+1)}},
\]
where each $f_i(x_n,t)(t+1)$ is obtained from Equation \ref{eq:bayes}. While we do get rid of the normalizing constant, the sum in the denominator prevents us from separating the logarithm as sums. 

The exception is the symmetric case, but only when we are updating the opinion $\nu_{in}$, where $n$ is exactly the choice of the neighbor $j$ and all other choices are equivalent. In that case, for every specific $n$, all values of $\mathbb{L}_{mn}$, when $m\neq n$, are equal to the same value $l=\frac{1-L_{nn}}{M-1}$. In this case, the terms in the denominator, $\sum_{m\neq n}f_i(a_n,t)\mathbb{L}_{mn}$ can be factored as we have  
\[
\sum_{n\neq m}f_i(a_n,t)\mathbb{L}_{mn}=l\sum_{m\neq n}f_i(a_m,t)=l (1-f_i(a_n,t)).
\]
Therefore, we obtain, for the opinion associated with the same choice $a_m$ as the neighbor,
\begin{equation}\label{eq:symmetricupdate}
\nu_{im}(t+1)=\nu_{im}(t)+\ln{\left[\frac{(M-1)L_{mm}}{1-L_{mm}} \right]}.
\end{equation}
%and, for the opinion about the other possible results, when $o\neq m$,
%\[
%\nu_{io}(t+1)=\nu_{io}(t)+                 \ln{\left[\frac{(M-1)L_{mm}}{1-L_{mm}} \right]}.
%\]
Unfortunately, as we need to update $\nu_{in}$ for all values of $n$ when an interaction happens, this is not a viable choice.

A distinct alternative, not as elegant as working directly with the values $\nu_{in}$ would have been, was proposed for the case where there are exactly three choices \cite{martins10b}. Instead of using the ratio between the probability of one choice and its negation, the odds between two actual choices can be used. That is, we can define (omitting the suffix identifying agent $i$) $\nu_{12}=\ln (f(1) /f(2))$, $\nu_{23}=\ln (f(2) /f(3))$, and $\nu_{31}=\ln (f(3) /f(1))$. As, for $M=3$, only two of the three probabilities are really independent, $\nu_{31}$ can be trivially obtained from $\nu_{31}=-\nu_{13}=-(\nu_{12}+\nu_{23})$. In the general case with $M$ choices, we have $M$ of those pairwise log-odds variables where $M-1$ of them are independent. The state of the system can be then described by the set of variables
\begin{equation}\label{eq:defnu}
\nu_{q(q+1)}= \ln{\frac{f(q)}{f(q+1)}},
\end{equation}
where $q$ assumes values in the range $1,\cdots,M-1$. And, of course, $\nu_{M1} = -\sum_{q=1}^{M-1} \nu_{q(q+1)}$. It is worth noticing that $\nu_{q(q+1)}$ contains all the information about the relative probability (odds) between choices $q$ and $q-1$. The odds that favor $q$ over $q-1$ are trivially $\exp(\nu_{q(q+1)})$. But the actual probability values for $q$ and $q-1$ are not obtainable without the complete set  of $\nu$s.

When we calculate the update rule, the renormalization terms cancel out and the final result is a simple multiplication. Once more, by working with log-odds we obtain simple additive dynamics for the variables $\nu_{q(q+1)}$. If agent $j$ is observed to prefer choice $m$, we have, for every $q=1, \ldots , M$
\[
\nu_{q(q+1)}(t+1)=\ln{\frac{f(q,t+1)}{f((q+1),(t+1))}}=\ln{\frac{f(q,t)\mathbb{L}_{mq}}{f((q+1),t)\mathbb{L}_{m(q+1)}}},	
\]
and, therefore, we get the general additive equation
\begin{equation}\label{eq:bayesmanychoices}
\nu_{q(q+1)}(t+1)=\nu_{q(q+1)}(t)+\ln \left[ \frac{\mathbb{L}_{mq}}{\mathbb{L}_{m(q+1)}} \right].
\end{equation}

Obtaining the $M$ probability values $f(q)$ can be done by solving the system of $M-1$ equations $\nu_{q(q+1)}=\ln (f(q) /f(q+1))$ plus the normalization condition. 

If solving that system was needed to check which value of $f(n)$ was the largest one and, therefore, the choice of the agent, solving the system could make actual implementations of the model more demanding computationally. Luckily, this is not needed, despite the fact that the variables $\nu_{q(q+1)}$ only compare pairs of choices. The procedure to find $i$ choice is simple. Take the first pair,  $\nu_{12}$. If it is a positive number, 1 is preferred to 2; otherwise 2 is preferred. We can drop one of the two in our search for the biggest probability and pick the one with larger probability, even without calculating the probabilities. If 2 is preferred to 1, check it against 3, using  $\nu_{23}$. If 1 is preferred, however, we actually need  $\nu_{13}$, for the comparison and this is not directly calculated by the update rule. However, it is trivial to check that $\nu_{13}=\nu_{12}+\nu_{23}$ and we can easily compare 1 and 3 and we will have the choice among the first three options that has the largest probability. We just have to repeat this procedure until we have checked all $M$ possible choices.

\subsection{Choosing an update matrix $\mathbb{L}_{mm}$}

In the general case, most terms of the matrix $\mathbb{L}_{mm}$ are independent of one another. The only rule they must obey in the general case is the trivial restriction that $\sum_m \mathbb{L}_{mn}=1$ for each possibility $n$. That is nothing more than stating that, if $n$ is the best choice, the chances agent $j$ will pick each possible preference must add to 1. That means we have $M^2-M$ independent terms in the general case and the choice of matrix $\mathbb{L}_{mm}$ must be performed carefully in order to reflect the scenario one is simulating. That multitude of parameters for large $M$ reflects the fact that evidence in favor of one choice might influence the other choices relative importance in different ways depending on the scenario.

Fortunately, for the most symmetrical case, the number of independent terms drops drastically to $1$, regardless of $M$. That case, studied bellow, corresponds to the situation where all choices are, in principle, identical and independent. To illustrate and better understand the more general case, I will also discuss the problem of choosing $\mathbb{L}_{mm}$ where each choice represent a position over a one-dimensional issue. Simulations, however, will be limited to the symmetric case.

\subsubsection{Symmetric Choices}

In the general case, the only constraint on the likelihood matrix  $\mathbb{L}_{mn}$ is the fact that probabilities conditional to the same possibility must add to one. Luckily, a common and interesting case corresponds to the symmetrical situation where all choices are equivalent. The choice of a sport team to cheer for is an example of that problem. If we assume complete symmetry between all options, we must have that $\mathbb{L}_{mm}=\mathbb{L}_{nn}$ for all values of $m$ and $n$, as well as $\mathbb{L}_{mq}=\mathbb{L}_{nq}$, for every $m\neq q$ and $n\neq q$. As consequence, the matrix $\mathbb{L}_{mn}$ will have only one independent value $l$, that we can choose as the diagonal term, where $\mathbb{L}_{mm}=l$ for every $m$. As the other choices in each line are all equivalent, they must have the same value and since probabilities for each possibility must add to one we have that $\mathbb{L}_{mn}=\frac{1-l}{M-1}$ for every $m \neq n$. Unless agents were contrarians \cite{galam04,delalamaetal05a,martinskuba09a,hongstrogatz11a}, they should also expect that, when $n$ is the best choice, it should be observed more often than the other alternatives and, therefore, we must choose values of $l$ that respect the condition $l>\frac{1-l}{M-1}$.

In this case, Equation \ref{eq:bayesmanychoices} can be written, when $m$ is the observed choice of agent $j$, as
\[
\nu_{q(q+1)}(t+1)=\nu_{q(q+1)}(t)+\ln \left[ \frac{\frac{1-l}{M-1}}{\frac{1-l}{M-1}} \right] = \nu_{q(q+1)}(t),
\]
when $q\neq m$ and $q+1 \neq m$. When one of those inequalities fail, we have
\[
\nu_{q(q+1)}(t+1)=\nu_{q(q+1)}(t)+\ln \left[ \frac{l}{\frac{1-l}{M-1}} \right] = \nu_{q(q+1)}(t) + \lambda
\]
for $q = m$, where $\lambda = \ln \left[  \frac{l(M-1)}{1-l}  \right] >0 $, and
\[
\nu_{q(q+1)}(t+1)=\nu_{q(q+1)}(t)+\ln \left[ \frac{\frac{1-l}{M-1}}{l} \right] = \nu_{q(q+1)}(t) - \lambda
\]
for and $q+1 = m$. That is, only the two log-odds variables directly related to $m$ need to be updated. Interestingly, we can define a renormalized log-odds as $\nu_{qm}^{*}= \nu_{qm}/\lambda$, so that the update rule becomes, substituting the values of $q$ for the only cases where there is change,
\begin{eqnarray}\label{eq:bayesmanychoicessymm}
\nu_{(m-1)m}^{*}(t+1) = \nu_{(m-1)m}^{*}(t)-1\nonumber\\
\nu_{m(m+1)}^{*}(t+1) = \nu_{m(m+1)}^{*}(t)+1
\end{eqnarray}
The meaning of this update rule is trivial. For the pairs that do not involve $m$, no change happens. For the two variables $\nu$ that relate to $m$, the odds in favor of $m$ are increased in both cases. Equations \ref{eq:bayesmanychoicessymm} mean $m-1$ becomes less likely than $m$ and $m$ becomes more likely than $m+1$.

One interesting characteristic of this case, shared with the original CODA model, is that, after renormalization, we obtain a model where even the single independent value $l$ of the matrix $\mathbb{L}_{mn}$ is irrelevant to the dynamics. We can write the whole algorithm and perform the simulation with not a single use for the value of $l$. The renormalized variables, in this case, simply tell us how many interactions in the right direction it would take to reverse the preference between choices $m$ and $m-1$. If one want to recover actual probability values for the choices, then $l$ is still needed. But that is its only influence in the model. Choices do not depend on its actual value.

\subsubsection{A one-dimensional issue}

Another interesting case happens when choices correspond to positions on a issue that allows for more than two simple choices. Quite often, those positions can be reasonable represented in a geometrical space \cite{Black1948,Downs1957,Miller2015}. For example, people might identify themselves along a left-right axis as only left or right. But it is also possible to observe more options. Centrists are one possibility. Extreme left and extreme right are more possible cases. For some applications, therefore, even a simple one dimensional choice might correspond to 3, 5 or any other number of choices. Of course, those choices can be modeled as a continuous variable \cite{deffuantetal00,Maciel2018}. 
But, depending on the problem, we might want to keep the discrete choice as a better way to describe the situation.

Assuming the choices are equivalent to positions on a line implies that some inequalities must be obeyed by the elements in $\mathbb{L}_{mn}$. Take, for example, $M=4$ and, without loss of generality, the values of $q$ are ordered according to the order in the line. For each possibility $n$, values of $m$ that are more distant from $n$ must correspond to smaller chances of observation given by $\mathbb{L}_{mn}$. As an example, the probability other agents will pick choices 1, 2, 3, or 4 if the best one were 1 must be in decreasing order. In that case, we must have $\mathbb{L}_{11}>\mathbb{L}_{12}>\mathbb{L}_{13}>\mathbb{L}_{14}$. That is, if the more extreme option 1 was the best, the chance a neighbor would choose it should be the largest one, as always. But, more than that, as 2, 3, and 4 are increasingly further from 1, their likelihood must decrease. Ordering for  different assumptions of which choice is best should also be included. For example, choice $1$ should be a more likely choice for agents if the actual best option is closer to 1. Notice that this corresponds to conditionals that are inverted when compared to the previous ordering. Here, we should also have $\mathbb{L}_{11}>\mathbb{L}_{21}>\mathbb{L}_{31}>\mathbb{L}_{41}$.

While those inequalities are how we translate the one-dimensional structure in the $\mathbb{L}_{mn}$ matrix, they do not actually decrease the number of parameters. That is the general case for the one-dimensional problem. But, if we do assume a few symmetries, we can still work with a few less parameters. The simplest symmetry we can introduce is a reflection symmetry around the central preference (or pair of preferences). That is, we assume the most extremes choices are equivalent as well as any two options that are equally distant from the two most extreme cases. For the example with four choices, we have no central choice, but we can still obey a reflection symmetry by making $\mathbb{L}_{m1}$ be the mirror of $\mathbb{L}_{m4}$, and the same with $\mathbb{L}_{m2}$  and $\mathbb{L}_{m3}$. That cuts the number of update parameters in half, in this example where $M=4$, from 12 to 6.

The question of comparing outer lines (or columns) of the matrix $\mathbb{L}_{mn}$ (lines 1 and 4 in the example) to the inner ones (2 and 3) is not as trivial as it might first appear, however. It might be reasonable to assume that all diagonal parameters $\mathbb{L}_{mm}$ should be the same in a symmetric problem. That was the case when we had 4 identical and independent options. But, in the current spatial description, it is clear that extreme choices are not the same as intermediate ones. Symmetry arguments do not work so well here, as those choices are not necessarily equivalent. The symmetry already imposed in the previous paragraph implies that $\mathbb{L}_{11}=\mathbb{L}_{44}$ and $\mathbb{L}_{22}=\mathbb{L}_{33}$. But it does not imply that those two sets should also be equal. That can be the case in a specific implementation, thus reducing the number of free parameters in $\mathbb{L}_{mn}$ to 5. But there is no a priori reason to impose that condition.

Furthermore, one might consider that the elements in any inner line or column should be symmetric. That is, for $M=4$ choices, we could have, for example, $\mathbb{L}_{21}=\mathbb{L}_{23}$. Or, in plain terms, as 1 and 3 are equally distant from 2, the chance to observe them when 2 is the actual best choice should be the same. That might seem a reasonable assumption but, once more, it does not follow from the reflection symmetry. It is just as reasonable to consider that, if a moderate opinion is the best choice, the other moderate opinion would be more likely than the extreme one. Or to consider that if a moderate position to one side is the best choice, the extreme position at the same side should be more likely than the moderate opposite one. All those assumptions can be defended. They correspond to a choice of metric in the spatial model and that is a much stronger assumption than pure reflection symmetry. The exception is, when we have an odd number of possible choices, we will have an actual central one. For the line corresponding to that choice, reflection symmetry does apply. If the center is the best choice, moderate left and moderate right should be treated equally. The same goes for extreme right and extreme left.

Here I will not deal with the problem of defining proper metrics. That is beyond the scope of the paper and a full study of the one-dimensional problem will be left for the future. %Instead, in the simulation section, I will just choose reasonable sets of parameters to test how the system evolves. The purpose is to provide a different case to compare with the results of the more common symmetrical case. Only a few cases will be studied and a throughout exploration of choices on a spatial setting will be left for a future paper.

%\begin{equation}\label{eq:bayesmanychoicessymm}
%\begin{array}{ccc}
%\nu_{(m-1)m}^{*}(t+1) &=& $\nu_{(m-1)m}^{*}(t)+1\nonumber\\
%\nu_{m(m+1)}^{*}(t+1) &=& $\nu_{m(m+1)}^{*}(t)-1
%\end{array}
%\end{equation}

%\begin{figure}
%\centering
%\begin{tabular}{cc}
%\epsfig{file=self3_propa03inflexa04.eps,width=0.45\linewidth,clip=} & 
%\epsfig{file=self3_propa03inflexa06.eps,width=0.45\linewidth,clip=} \\
%\epsfig{file=self3_propa03inflexa08.eps,width=0.45\linewidth,clip=} &
%\epsfig{file=self3_propa03inflexa095.eps,width=0.45\linewidth,clip=}
%\end{tabular}
%\caption{Distribution of opinions for differentr amounts of inflexibles in favor of $A$. All simulations started with just a minority supporting $A$, $p_A = 30\%$ and correspond to groups of size 3, where the agents are influenced by their own opinion. The upper left distribution corresponds to $I_A=0.4$; the upper right, to $I_A=0.6$; lower left, $I_A=0.8$; and lower right to $I_A=0.95$.}\label{fig:self3_propa03}
%\end{figure}

\section{Simulation results}

To test how the agent choices and opinions evolve in the $M$ choices case, the model was implemented using the R software environment \cite{Rsoftware}. In every scenario, agents where arranged on a non-periodic non-directed square lattice that defined their neighborhood, that is, who they might observe. To represent more or less connected situations, simulations were run for agents connect to all their first neighbors, as well as connections to second, and also third level neighbors. In other words, agents are connected up to the $c$ level. Networks were implemented using the igraph package \cite{Csardi2006}.

\subsection{Symmetrical case}

A series of simulations were run for the symmetrical choices case. They were all implemented on a square $50 \times 50$ non-periodic lattice network where each node corresponds to an agent and the edges define the neighborhoods of each agent. Agents were connected with the nearest neighbors up to the $c$ level, that is, when $c=1$, all agents were connected to all their first neighbors and nobody else. The cases $c =1, 2, 3$ were studied. In order to observe the influence of the number of possible choices $M$ on the opinions of agents, simulations were run for the cases $M=4,6,8,10$. Each combination of the parameters $c$ and $M$ were run 20 times and the average behavior over several runs was observed.

\begin{figure}
\centering
\begin{tabular}{c}
\epsfig{file=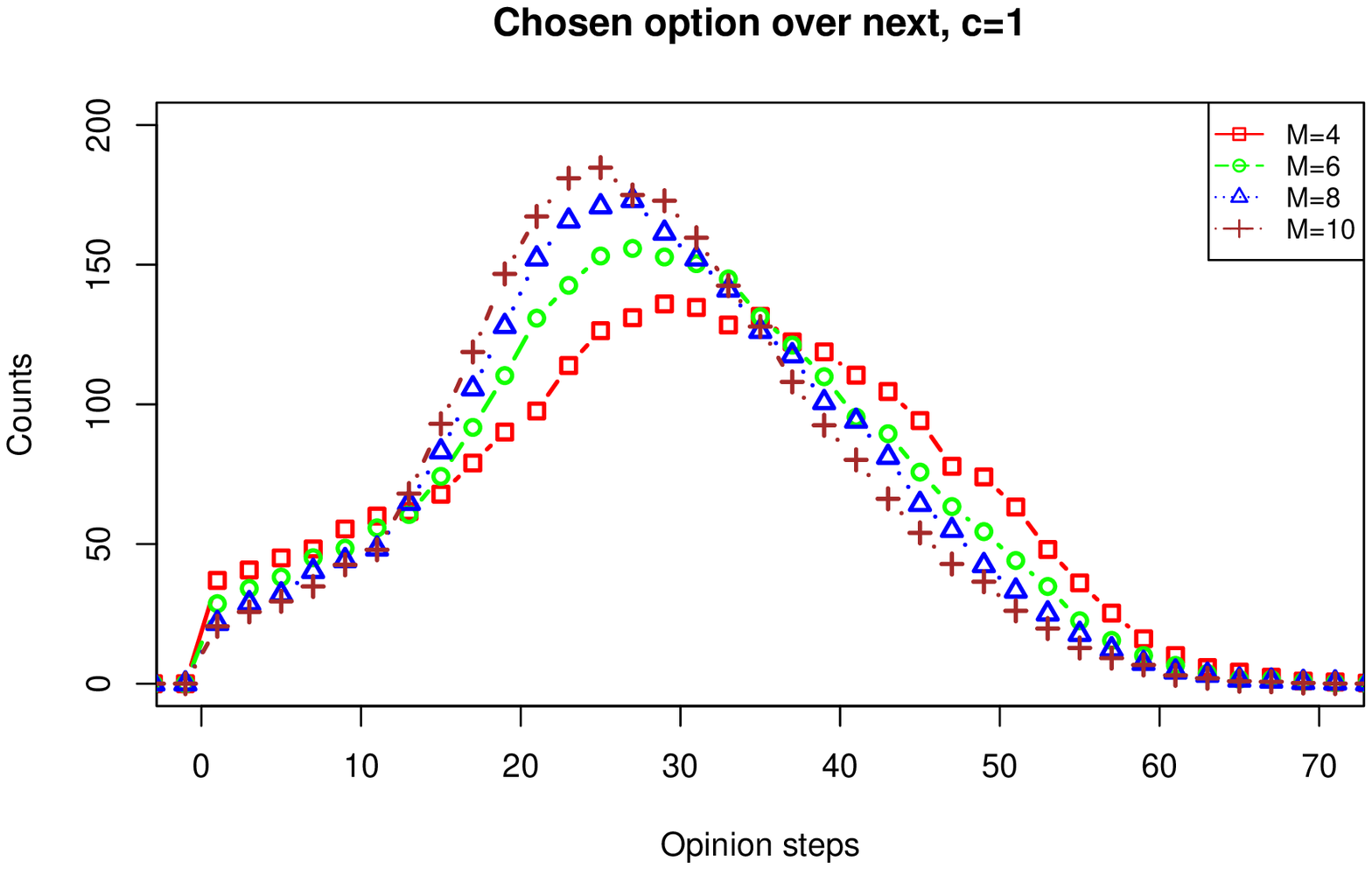,width=0.70\linewidth,clip=}\\
\epsfig{file=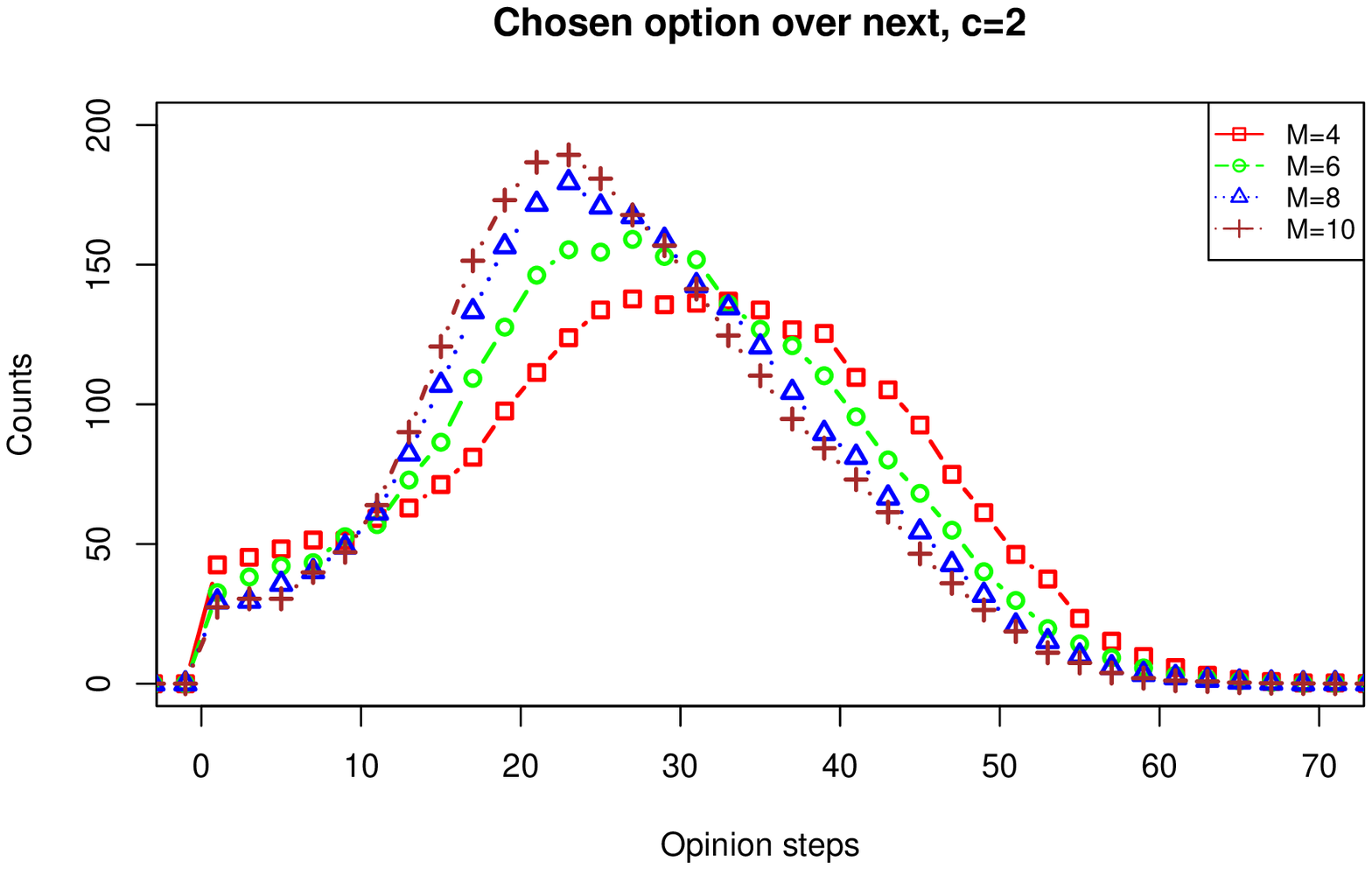,width=0.70\linewidth,clip=}\\
\epsfig{file=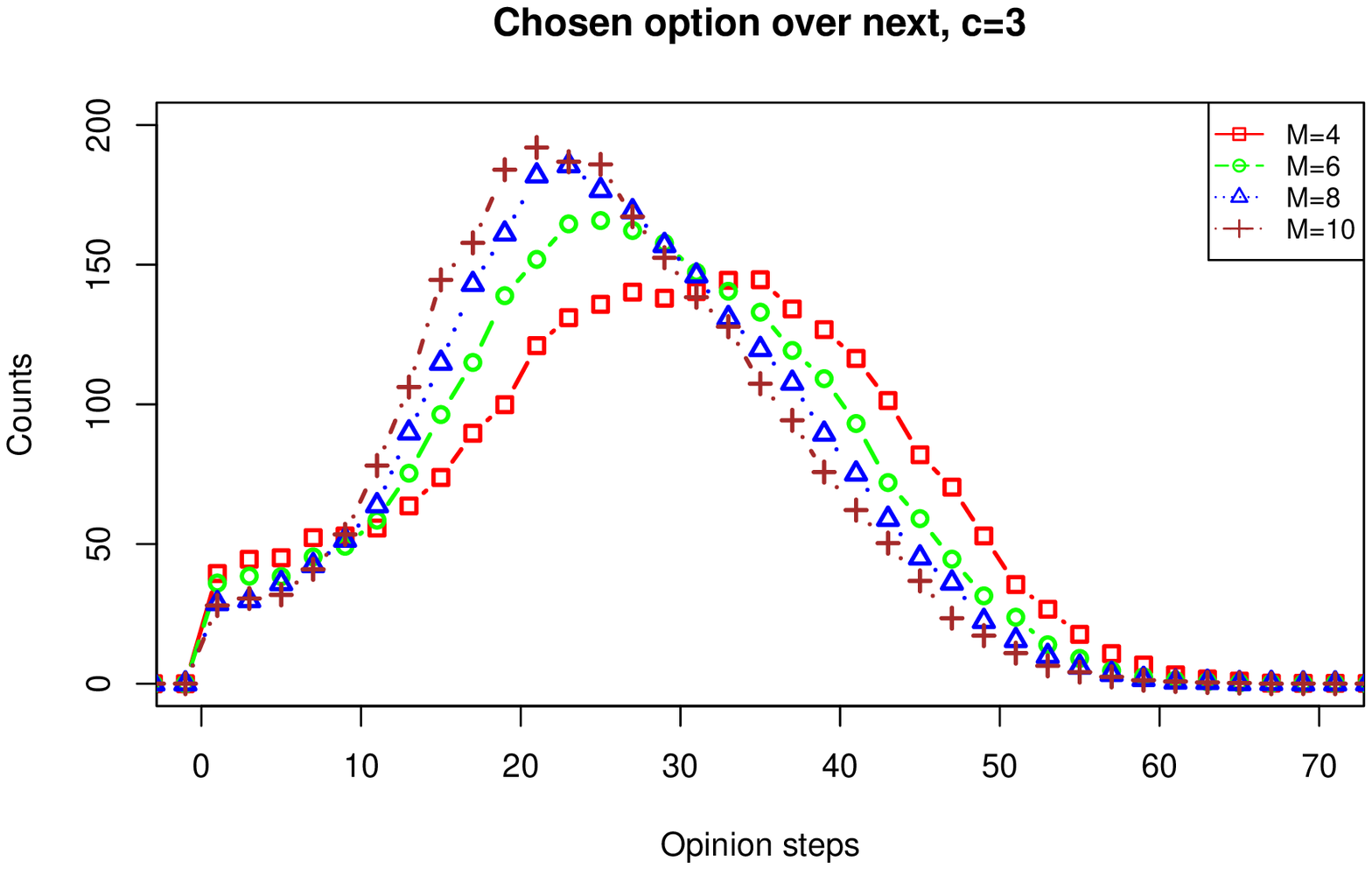,width=0.70\linewidth,clip=}
\end{tabular}
\caption{Distribution of opinions for different numbers of choices, $M$. The distribution for $\nu_{q_o(q_o +1)}$ is shown in all three graphics. {\it Top:} First neighbors connections ($c=1$). {\it Middle:} Second neighbors connections ($c=1$).{\it Bottom:} Third neighbors connections ($c=3$).
	}\label{fig:mainnext}
\end{figure}

Initial conditions were drawn randomly. For each agent $i$, an initial weight $w_m$ for each possible choice $m$ was randomly drawn between 0 and 1, to be made proportional to the probability agent $i$ assigned to choice $m$. From that, $\nu_{q(q+1)}$ for each $i$ was calculated simply by estimating $\nu_{q(q+1)}=\ln (w_q /w_{q+1})$.

Each interaction is done by picking one agent $i$ who will update its opinion randomly. A second agent $j$ is drawn from the set of those connected to $i$ and agent $i$ observes $j$ choice $m$. From that, $i$ updates its opinion variables $\nu_{q(q+1)}$ according to Equations~\ref{eq:bayesmanychoicessymm}. Agent $i$ new choice is calculated based on the updated $\nu_{q(q+1)}(t+1)$s. This procedure is repeated until, in average, each agent has had its opinion updated $T$ times. In all following implementations, $T=50$.

\begin{figure}
	\centering
	\begin{tabular}{c}
		\epsfig{file=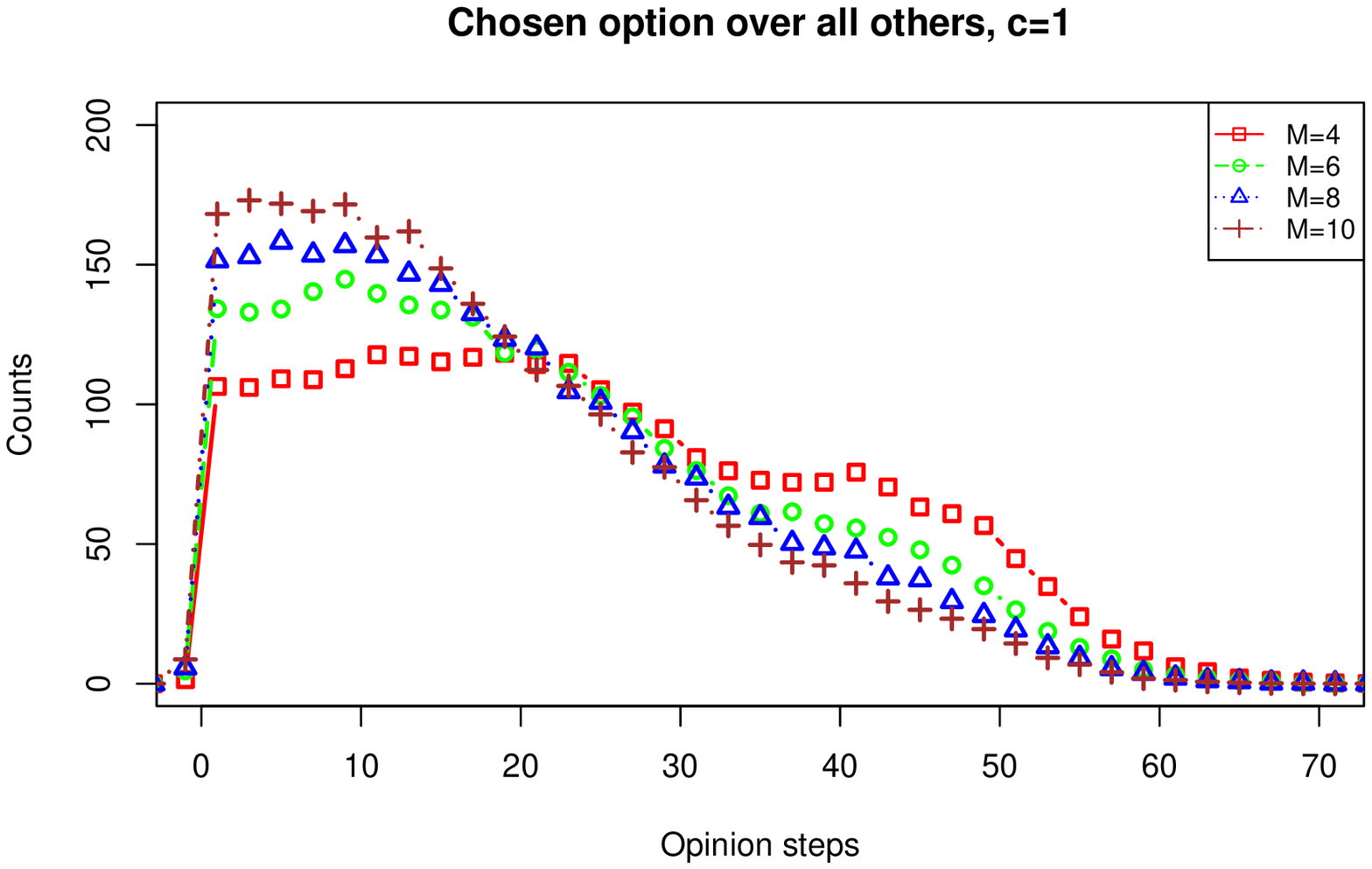,width=0.65\linewidth,clip=}\\
		\epsfig{file=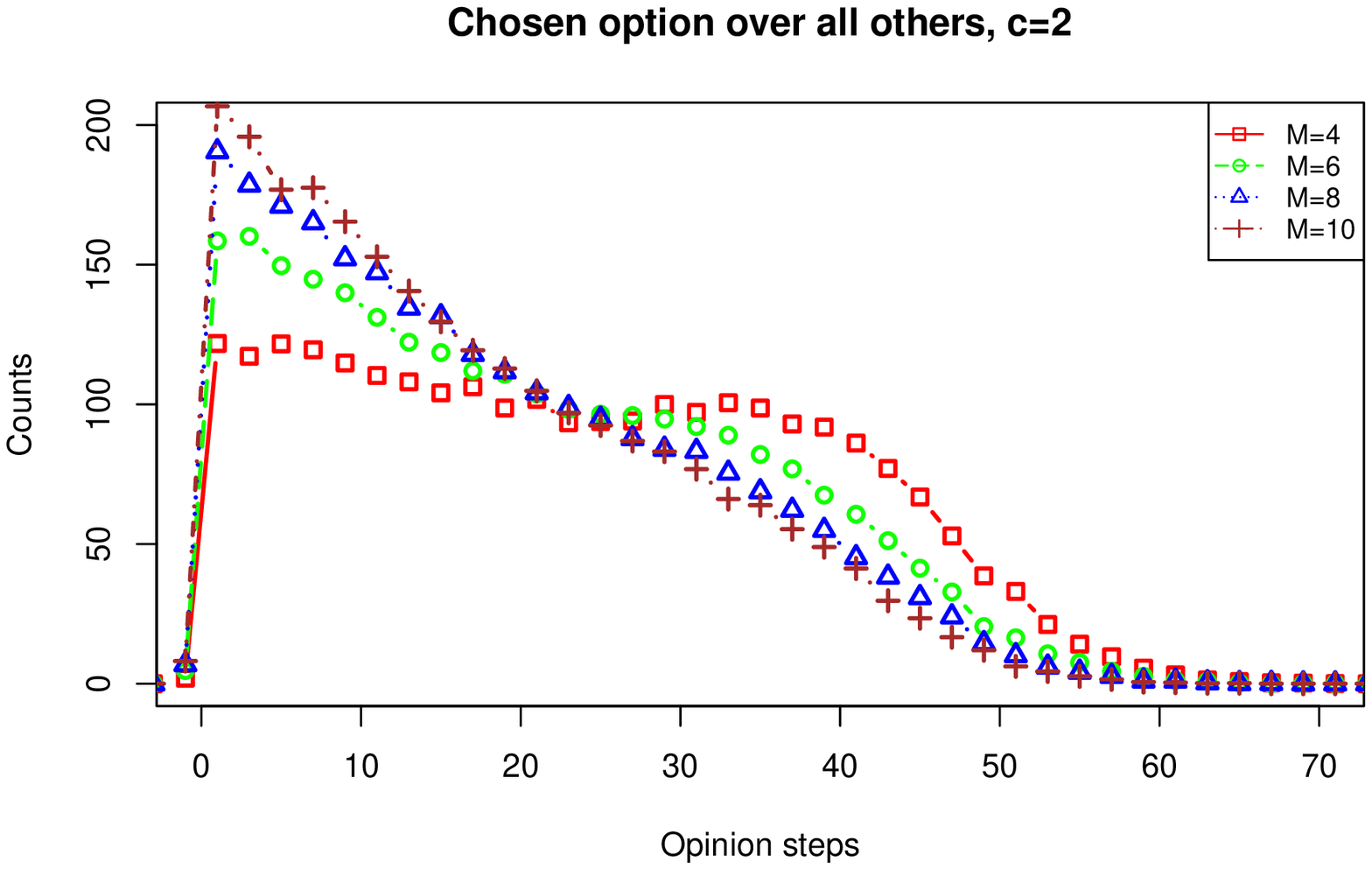,width=0.65\linewidth,clip=}\\
		\epsfig{file=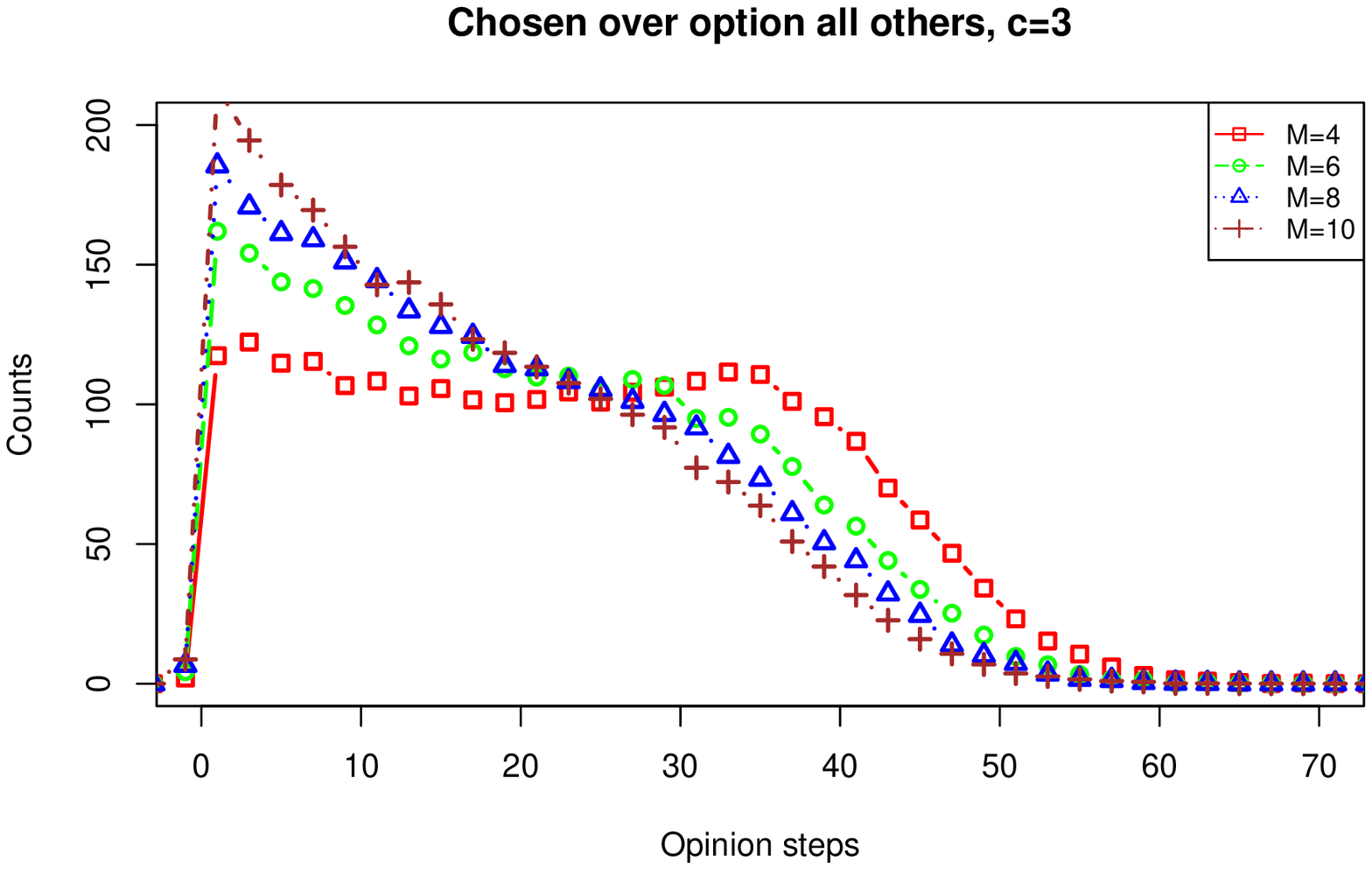,width=0.65\linewidth,clip=}
	\end{tabular}
	\caption{Distribution of opinions for different numbers of choices, $M$. The distribution for $\nu_{q_o}$ is shown in all three graphics. {\it Top:} First neighbors connections ($c=1$). {\it Middle:} Second neighbors connections ($c=1$).{\it Bottom:} Third neighbors connections ($c=3$).
	}\label{fig:mainstep}
\end{figure}

Measuring strength of opinion when there are many choices is not as straight-forward as when there are only two choices where the probability in favor of a choice is the same as the probability against the competing one. The simplifying variables $\nu_{q(q+1)}$ only provide pairwise comparisons. They tell us how much more (or less) likely is choice $q$ when compared to $q+1$, but they tell us nothing about $q$ alone or about every other choices. If a certain $q_o$ is the chosen option of agent $i$, $\nu_{q_o(q_o +1)}$ does tell us how much more probable the chosen option is than the next option. That next option, however, might just equally be the second preferred option or the least preferred one (or anything in between). That means that, while there is some interesting information on the distribution of values for $\nu_{q_o(q_o +1)}$, those distributions must be interpreted carefully. They only tell us how much more likely the best option $q_o$ is when compared to a random other choice.  

As an example, the distributions of $\nu_{q_o(q_o +1)}$ can be seen in Figure~\ref{fig:mainnext}. It is clear all three values of $c$ show the same basic behavior, with only very small changes in the curves that can be attributed to random variation. Strong opinions appear in all cases, with the peaks of most likely opinion strengths situated in the $20-30$ steps range. That means it would take 20-30 interactions with opposing views to change the preference of those agents for the option $q_o$ over the next one, $q_o +1$. Even cases around and above 50 are observed, basically suggesting that, for an important fraction of the agents, the majority of their interactions happened with neighbors who agreed and reinforced their positions. That was particularly true for smaller values of $M$, when the distributions had more important spreads to more extreme opinions. Still, even for $M=10$, very extreme opinions were observed. 

\begin{figure}
	\centering
	\begin{tabular}{ccc}
		\epsfig{file=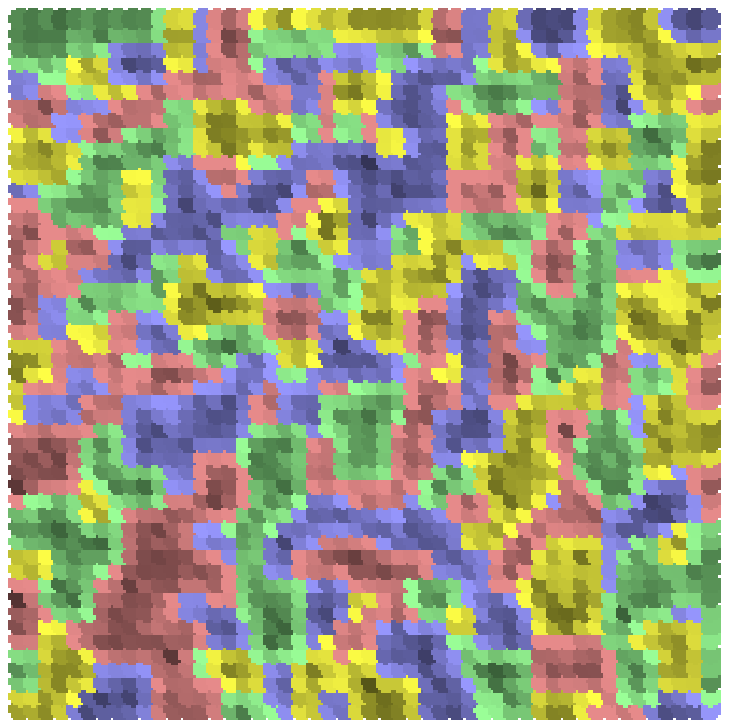,width=0.35\linewidth,clip=}&
		\epsfig{file=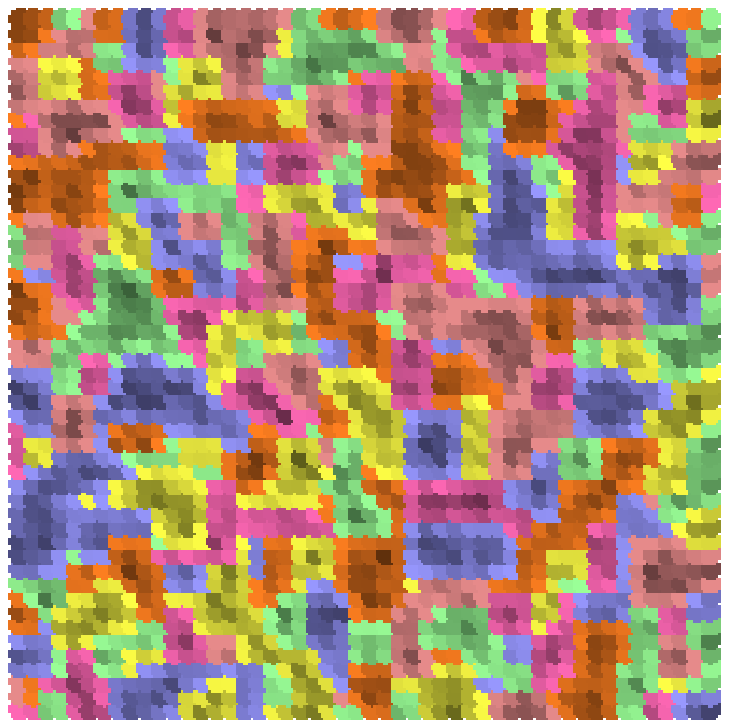,width=0.35\linewidth,clip=}&
		\epsfig{file=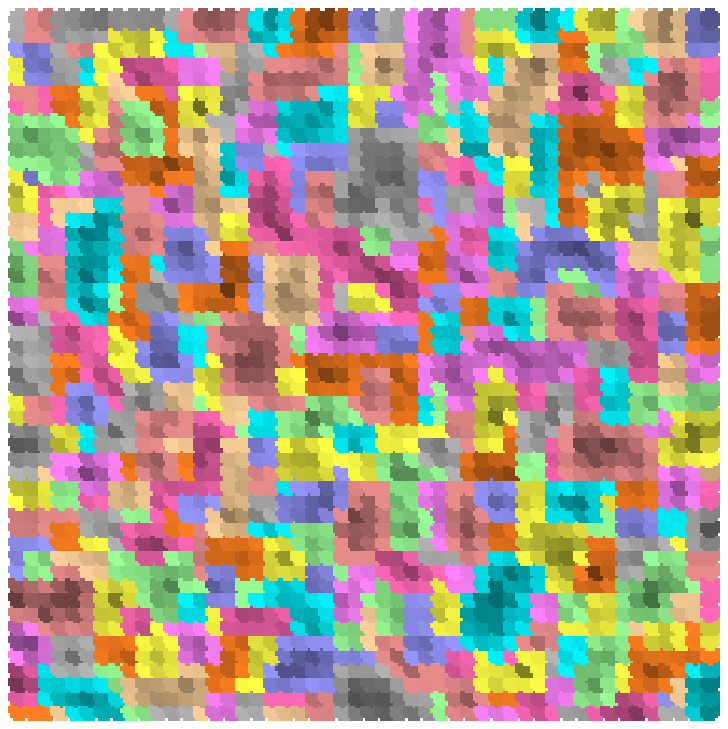,width=0.35\linewidth,clip=}\\
		\epsfig{file=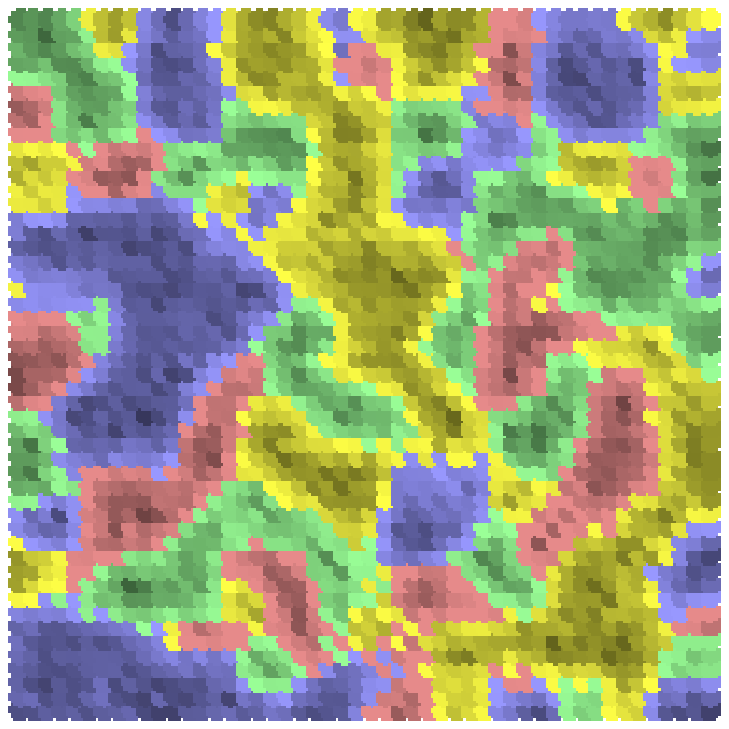,width=0.35\linewidth,clip=}&
		\epsfig{file=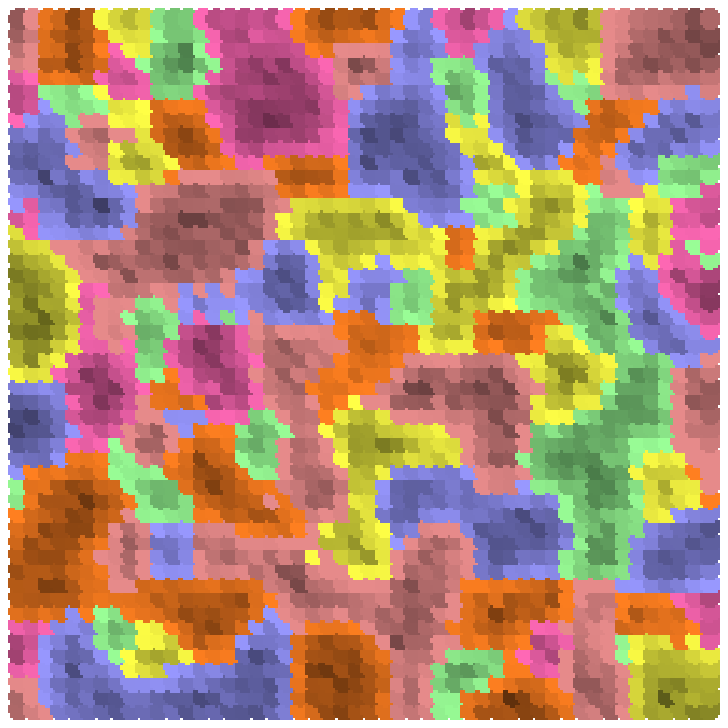,width=0.35\linewidth,clip=}&
		\epsfig{file=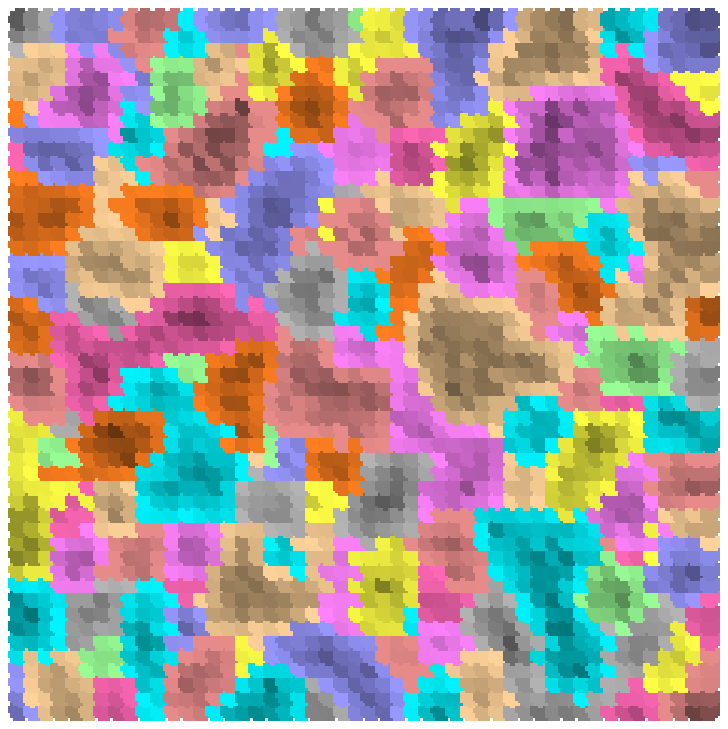,width=0.35\linewidth,clip=}\\
		\epsfig{file=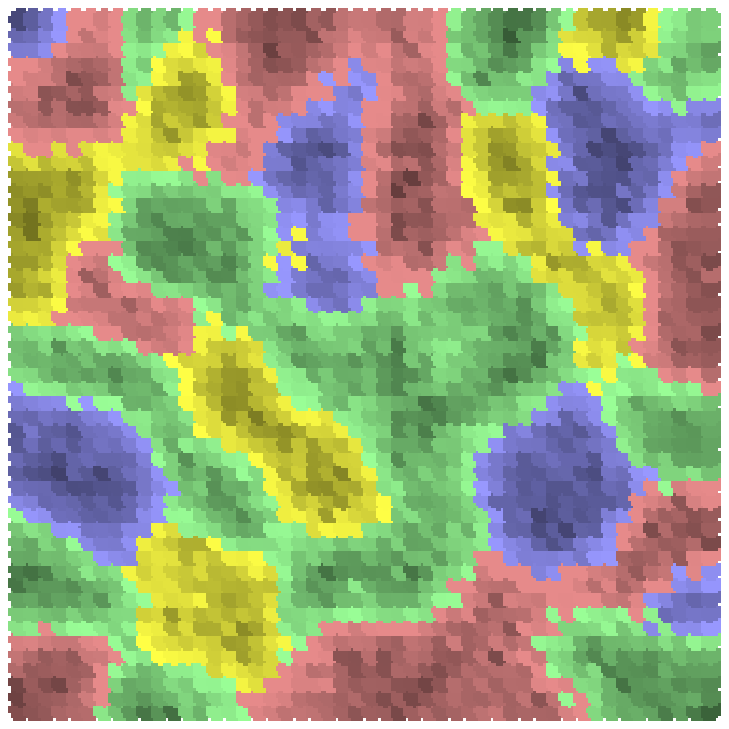,width=0.35\linewidth,clip=}&
		\epsfig{file=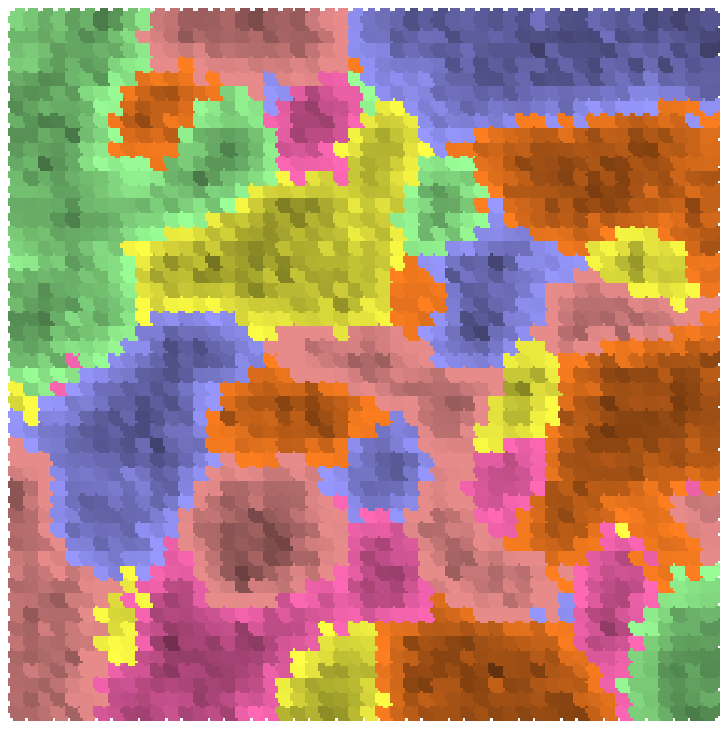,width=0.35\linewidth,clip=}&
		\epsfig{file=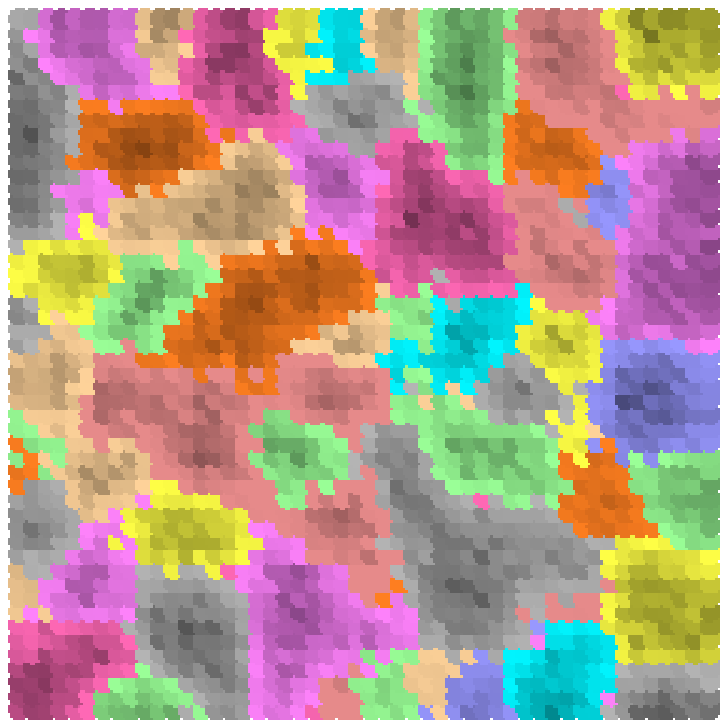,width=0.35\linewidth,clip=}
	\end{tabular}
	\caption{Final configurations after and average of 50 interactions per agent, for one example run of different values of $c$ and $M$. Different choices are represented with specific colors and color tones represent strength of opinion, with lighter versions corresponding to weaker opinions and darker ones corresponding to more extreme views. {\it Top line:} First neighbors connections ($c=1$). {\it Middle:} Second neighbors connections ($c=1$).{\it Bottom:} Third neighbors connections ($c=3$). {\it First column:} $M=4$. {\it Second column:} $M=6$. {\it Third column:} $M=10$. 
	}\label{fig:configs}
\end{figure}

To understand how extreme the opinions in favor of the preferred choice actually are, $\nu_{q_c(q_c +1)}$ is not the best measure, however. To answer that question more correctly, we must separate the impact of each choice. The values of the probabilities $f(q)$ can be easily obtained by inverting Equation~\ref{eq:defnu}. To obtain number of steps once more, we can estimate the log-odds of the best option $q_c$ against all others added, or, $\nu_{q_o} = \frac{f(o)}{1-f(o)}$. Solving the equations we get
\begin{equation}
\label{eq:nuqcnuqq1}
   \begin{split}
   \nu_{q_o} = \left[ e^{-\nu_{q_o(q_o +1)}} + e^{-(\nu_{q_o(q_o +1)} + \nu_{(q_{o} + 1)(q_o +2)} )} + \ldots + \right. \\ 
	 \left. + e^{-(\nu_{q_o(q_o +1)} 
		 + \nu_{(q_{o}+1) (q_o +2)} + \ldots \nu_{(q_o - 2)(q_o - 1)} )} \right] ^{-1} ,
	\end{split}
\end{equation}

where the sum has $M-1$ terms and is meant to continue in a circular way, the terms jumping from $q=M$ back to $q=1$. As $\nu_{q_o}$ is a direct comparison of the best alternative against the sum of all others, its interpretation is much more direct. It should also be noted that, while $\nu_{q_o(q_o +1)}$ must be greater than zero (since $o$ is the chosen option, it must be preferred to $o+1$), the same is not true for $\nu_{q_o}$, except when $M=2$. We can have an option that is preferred to all others when compared individually, but still with a probability of being the best one smaller than 50\%.

The distributions for $\nu_{q_o}$  can be seen in Figure~\ref{fig:mainstep}. While there are a few observations of values of $\nu_{q_o} < 0$, corresponding to a probability smaller than 50\%, they are very rare. That means the vast majority of agents have interacted with neighbors that agree with their position more than 50\% of the times, suggesting the existence of very clear opinion domains. Interestingly, while changing the size of the neighborhood had little effect at the distributions of $\nu_{q_o(q_o +1)}$, the same is not true for $\nu_{q_o}$. While the distributions for $M=4$ still seem little affected when we observe different values of $c$, there is a noticeable difference in behavior for higher values of $M$ when we have only first neighbors ($c=1$). When $c=1$, we observe for all values of $M$ the existence of plateaus at the weaker opinion values. followed by slow decrease in frequency as opinions get more extreme.  For $M=4$, that plateau is observed for all values of $c$ and, in all cases, it extends up to more extreme opinions. But when $c=2$ or $c=3$ and $M$ is larger then 4, the plateau is replaced by a clear peak close to $\nu_{q_o}=0$. We can also see that,  for all values of $M$, it seems that the more connected network corresponds to a smaller proportion of the most extreme opinions. We can also observe that, as $M$ increases, the region of extreme opinions become less important. But it does not disappear and even as $M=10$, we still observe agents that think their choice is more than 50 steps away from the sum of all others.

\begin{figure}
	\centering
	\epsfig{file=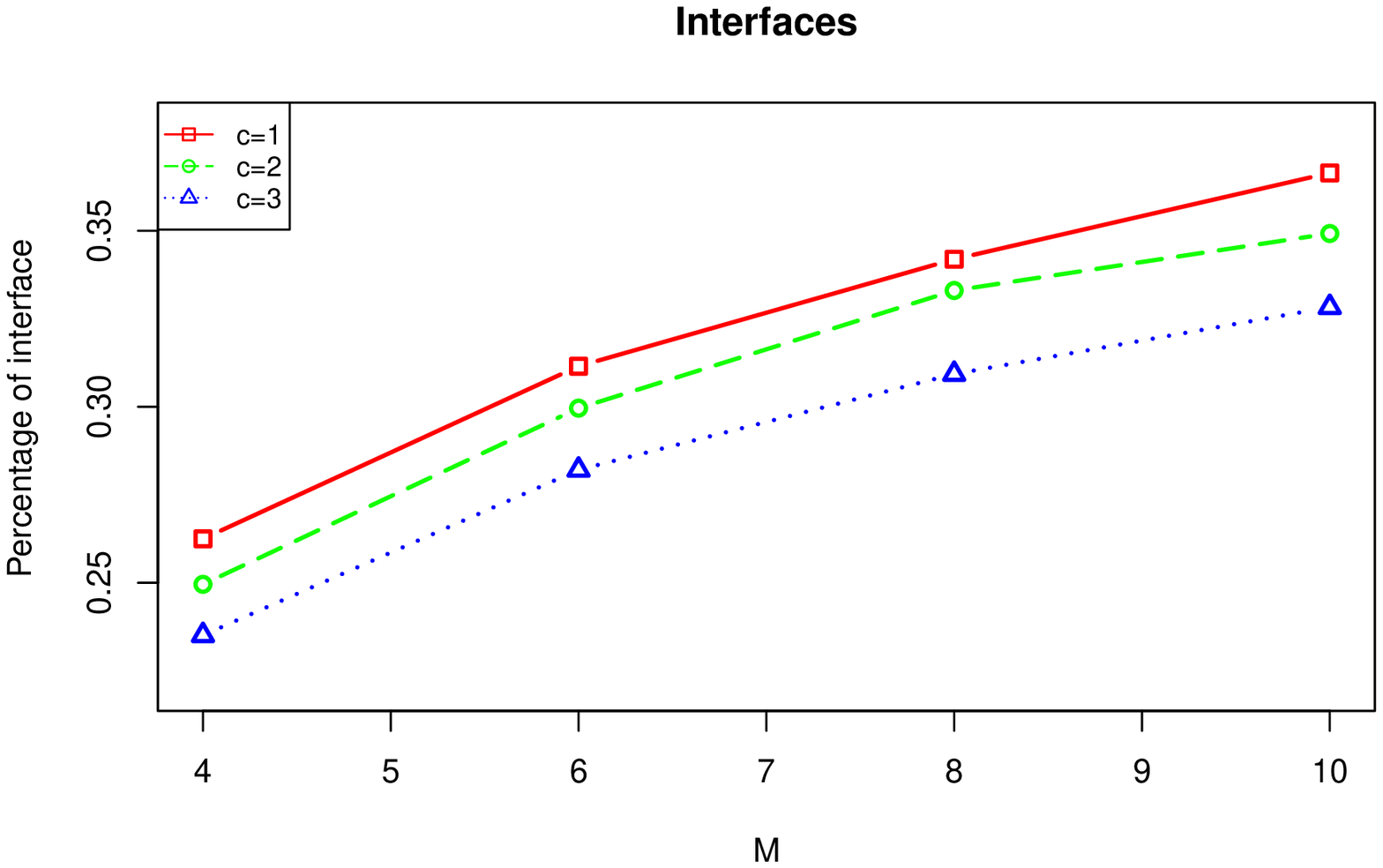,width=0.65\linewidth,clip=}
	\caption{Average proportion of edges that correspond to interfaces between choice domains.
	}\label{fig:interfaces}
\end{figure}

As in the original CODA model, extreme opinions close to the average number of interactions of each agent are a clear sign of domains. Inside a domain, an agent will only have its opinion reinforced and that explains its final extremist views. To visualize those domains, Figure~\ref{fig:configs} show typical final states  after an average of 50 interactions per agent, for different values of $c$ and $M$. Different choices are represented with specific colors and color tones represent strength of opinion, with lighter versions corresponding to weaker opinions and darker ones signaling more extreme views. The figures are organized so that each line correspond to a different size $c$ of neighborhood ($c=1,2,3$) and each column correspond to a different number of choices ($M=4,6,10$). It is easy to see that the domains become bigger with $c$, as it should be expected, since each agent influence can reach further. 

The size of domains seem to also change with $M$, as the graphics seem more crowded and with smaller regions, but visual inspection is not so clear now. To verify that is actually the case, Figure~\ref{fig:interfaces} shows the average over 20 runs of the proportion of interfaces between the domains. Here, the interface is defined as a network edge that links nodes with different choices. It is easy to notice that the proportion of interfaces decrease as $c$ gets bigger, meaning that larger values of $c$ do correspond to larger domains, as we had observed in the final configuration examples. The apparent observation that a larger number $M$ of choices correspond to smaller domains is also confirmed, as it is now clear that the proportion of interfaces does increase as $M$ does.

\section{Conclusion}

Here, I presented an extension of the CODA model to the case where you have any number $M$ of possible choices. A simple solution that leads to an additive model in the general case was found by using paired sequential log-odds between a choice and the next one. When all choices are independent and symmetric, we saw that the whole dynamics depends on a single update parameter. As in the original CODA model, if we are only interested in the dynamics of the opinions, we can renormalize the variables, obtaining a dimensionless model with no parameters. The original parameter is only needed if we want to assign actual probability values to the number of steps an agent is from flipping its opinion.

Extreme opinions were observed again in all runs, thanks to the local reinforcement that happens once domains form. For smaller values of choices, at $M=4$, a large plateau can be observed from weak to strong opinions, decaying only slowly. That general tendency to form a plateau seems to happen regardless of how many neighbors agents have. Similarly, smaller plateaus were also observed for all number of choices when only first neighbors interacted. As the network included more distant edges and $M$ was larger (6, 8, and 10), the plateau changed to a peak close to a probability of 50\% for the preferred choice ($\nu_{q_o}=0$). Both effects might be associated to the longer time it should take for rigid domains to form when there are more choices (larger $M$) and also because more distant neighbors have a larger chance to belong to a different domain, also contributing to more time being needed before domains stabilize. Indeed, as $c$ grows, we observed that the domains tend to be larger (less interfaces). Once those domains do get stable, the local dynamics for agents inside them will make those agents go to the extremes as every interaction becomes a reinforcing one. 

While only the common symmetrical case was used in the simulations, the model can be used in the general situation. The number of interaction parameters in that situation can be large, as can be expected of a general situation with no symmetries. Exploring all those possibilities, due to the number of parameters, is beyond the scope of this paper.

It is also interesting to notice that the variables that made modeling easier, turning the problem into sums even in the general situation, are not the natural ones to interpret. For measuring the strength of opinions, comparing each agent estimate of its choice against all others provides better data than comparing against a random next possibility. That makes a change of variables necessary once the simulation results are obtained, but only for questions about the strength of opinion. The evolution of choices and their domains can be easily described with the original model variables. 

%The diminishing in extremism as $M$ increases in the symmetrical case is weak. It is possible it might happen because, as there are more options, the agents might have, at first, before the domains get formed, a larger exposure to many choices. That might be simply an effect that, with more choices, the chances of reinforcement in the initial random situation are smaller.

%\section{Acknowledgements.} One of the authors (ACRM) would like to thank the Funda\c{c}\~ao de Amparo a Pesquisa do Estado de S\~ao Paulo (FAPESP) for the support to the work under grant 2009/08186-0.

%%%%%%%%%%%%%%%%%%%%%
%To be completed

\bibliographystyle{unsrt}
\bibliography{biblio}

\end{document}